\documentclass[trackchanges]{aastex631} 

\shorttitle{Flux rope in an RMHD Simulation}
\shortauthors{Wang et al.}
\newcommand{\rrr}[1]{{{}}}

\begin{document}

\title{Radiative Magnetohydrodynamic Simulation of the Confined Eruption of \\a Magnetic Flux Rope: Magnetic Structure and Plasma Thermodynamics}

\correspondingauthor{Feng Chen}
\email{chenfeng@nju.edu.cn}

\author{Can Wang}
\affiliation{School of Astronomy and Space Science, Nanjing University, Nanjing 210023, China}
\affiliation{Key Laboratory for Modern Astronomy and Astrophysics (Nanjing University), Ministry of Education, Nanjing 210023, China}

\author[0000-0002-1963-5319]{Feng Chen}
\affiliation{School of Astronomy and Space Science, Nanjing University, Nanjing 210023, China}
\affiliation{Key Laboratory for Modern Astronomy and Astrophysics (Nanjing University), Ministry of Education, Nanjing 210023, China}

\author{Mingde Ding}
\affiliation{School of Astronomy and Space Science, Nanjing University, Nanjing 210023, China}
\affiliation{Key Laboratory for Modern Astronomy and Astrophysics (Nanjing University), Ministry of Education, Nanjing 210023, China}

\author{Zekun Lu}
\affiliation{School of Astronomy and Space Science, Nanjing University, Nanjing 210023, China}
\affiliation{Key Laboratory for Modern Astronomy and Astrophysics (Nanjing University), Ministry of Education, Nanjing 210023, China}

\begin{abstract}
It is widely believed that magnetic flux ropes are the key structure of solar eruptions; however, their observable counterparts are not clear yet. We study a flare associated with flux rope eruption in a comprehensive radiative magnetohydrodynamic simulation of flare-productive active regions, especially focusing on the thermodynamic properties of the plasma involved in the eruption and their relation to the magnetic flux rope. The pre-existing flux rope, which carries cold and dense plasma, rises quasi-statically before the eruption onsets. During this stage, the flux rope does not show obvious signatures in extreme ultraviolet (EUV) emission. After the flare onset, a thin `current shell' is generated around the erupting flux rope. Moreover, a current sheet is formed under the flux rope, where two groups of magnetic arcades reconnect and create a group of post-flare loops. The plasma within the `current shell', current sheet, and post-flare loops are heated to more than 10 MK. The post-flare loops give rise to abundant soft X-ray emission. Meanwhile a majority of the plasma hosted in the flux rope is heated to around 1 MK, and the main body of the flux rope is manifested as a bright arch in cooler EUV passbands such as AIA 171 \AA~channel.
\end{abstract}


\section{Introduction} \label{sec:intro}
Solar flares and coronal mass ejections (CMEs) are the most violent solar activities that result from a rapid release of magnetic energy in the solar atmosphere. It is now widely believed that flares and CMEs are two different manifestations of the same magnetic process \citep{Zhang2001,Lin2004,Webb2012}, in which magnetic flux ropes act as the key ingredient involved in the eruption \citep{Nindos2015,Schmieder2015,Cheng2020}.

A magnetic flux rope is characterized by a group of magnetic field lines that coherently twist around a central axis, showing evidently a helical morphology that is clearly distinct from the ambient magnetic structures \citep{Liu2020}. Based on the observing wavelengths, a variety of observational manifestations have been proposed to be related to flux ropes, such as filaments/prominences \citep{Mackay2010,Ouyang2017}, cavities \citep{Regnier2011,Sarah2015}, hot channels \citep{Zhang2012,Cheng2013}, and sigmoids \citep{Green2007,McKenzie2008}. Moreover, quite some magnetohydrodynamic (MHD) simulations based on different codes \citep[e.g.][]{Kliem2013,Xia2014,Inoue2014,Guo2019,Fan2019,He2020} have been performed, which try to explain the roles of flux ropes in solar eruptions and their manifestations in observations. For example, the simulations by \citet{Amari2000,Amari2003} ascribed the initiation of two-ribbon flares and CMEs to twisted flux ropes, which are formed through flux emergence or convergence after photospheric shearing motions. \citet{Aulanier2010} analyzed the formation and eruption of a torus-unstable flux rope in a zero-$\beta$ MHD simulation, and found that the evolution of the current system relevant to the flux rope can explain the X-ray sigmoids in observations. More advances in the modeling and observations of flux ropes can be found in a number of recent reviews \citep[e.g.][]{Filippov2015,Cheng2017,Patsourakos2020,Liu2020}.

However, owing to the difficulty in measuring the magnetic field in the corona, the observational counterparts of flux ropes are still elusive. Most of the previous MHD simulations have adopted more or less simplifications in the energy equation governing the plasma thermodynamics. Except for the very few data-driven simulations, most simulations started from a simplified, prescribed magnetic structure. Radiative magnetohydrodynamic (RMHD) simulations with sophisticated energy equation allow a direct comparison between model synthesized observable and real observations. \citet{Cheung2019} showed that RMHD simulation can successfully reproduce many of observational properties in solar flares.

In this Letter, we investigate a flux rope eruption in an RMHD simulation of active-region-scale flux emergence from the convection zone to the corona \citep{Chen2021}. The simulation provides a comprehensive history of the fully self-consistent evolution of the magnetic field starting from the interior. This is combined with a sophisticated energy equation that provides thermal properties of plasma and synthetic observables that are quantitatively comparable with observations. Thus, this study helps to shed new light on the relation of proposed emission counterparts of flux ropes and the true magnetic structures that are not directly observable. 

The rest of the Letter is organized as follows. A brief description of the numerical simulation is given in Section \ref{sec:method}. Section \ref{sec:result} presents the analysis of the magnetic field, the thermal properties of the plasma and the corresponding observables. We summarize the results and conclude in Sect \ref{sec:conclusion}.

\section{Method}\label{sec:method}

The numerical simulation is conducted with the MURaM code\citep{Voegler+al:2005,Rempel:2017} that takes into account radiative transfer in the lower solar atmosphere, optically thin radiative loss in the transition region and corona, and anisotropic thermal conduction along magnetic field lines. 
The simulation domain covers a region of $x\times y\times z = $ 196.6 Mm $\times$ 196.6 Mm $\times$ 122.9 Mm, and the lower boundary is placed at about 10 Mm below the photosphere.  The domain is resolved by 1024 $\times$ 1024 $\times$ 1920 grid points, yielding a resolution of 192 km in the horizontal direction and 64 km vertically. The full evolution of 48 hr covers the emergence of magnetic flux bundles from the convection zone to the corona. Complex active regions are formed in the photosphere and give rise to diverse solar activities including more than 50 flares of C class or above. Comprehensive descriptions on the simulation setup and general results are presented in \citet{Chen2021}. 

Here, we study a C8.5 flare, which is the second largest eruption happening in the simulation and focus on the region (580 $\times$ 400 $\times$ 1620 grid points) of the flare and a time period of approximately 1700 s starting from 21$\rm^{h}$36$\rm^{m}$55$\rm^{s}$ ($t_{0}$) in the simulation time domain. All the time instants mentioned hereafter are relative to $t_{0}$.

\section{Results} \label{sec:result}

\subsection{Synthesized Observables of the Event}

\begin{figure*}
\centering
\includegraphics[width=\textwidth]{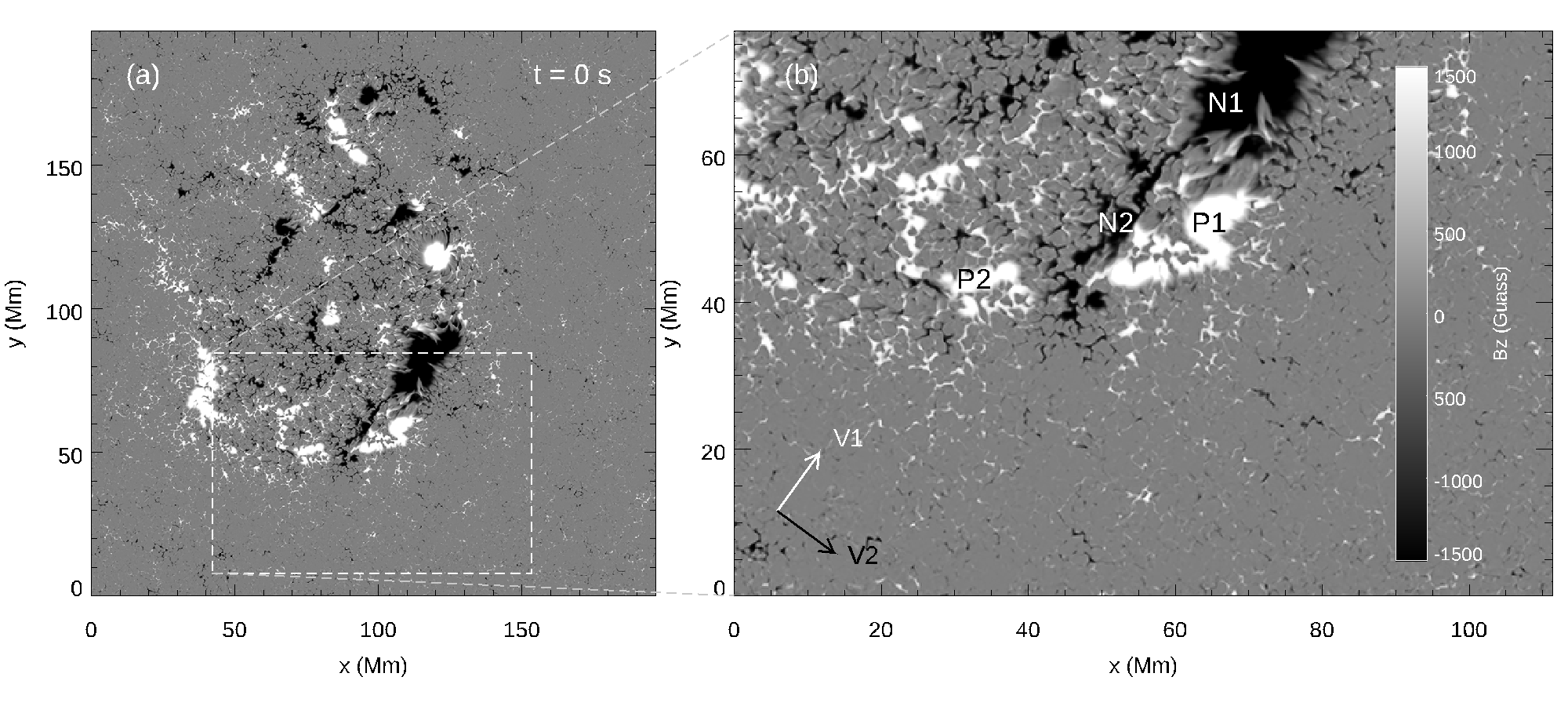} 
\caption{Photospheric magnetogram of the simulation at $t_{0}$. (a) The magnetogram of the full simulation domain at $t_{0}$. The region of interest is marked by the white dotted box. (b) A zoomed-in view of the region of interest shown in panel (a). There are two bipoles that are labeled as N1, P1 and N2, P2. The white arrow V1 and the black arrow V2 indicate the two viewing directions for synthesizing the AIA images. Note that the coordinates are relative ones that only measure the size of the region.}\label{fig:magneto}
\end{figure*}

The photospheric line-of-sight magnetic field of the full simulation domain at $t_{0}$ is shown in Figure \ref{fig:magneto}(a). The entire region consists of several strong sunspots, with the white box indicating the region of interest for this study. This region covers two bipoles, which are marked as N1, P1 and N2, P2 in a zoom-in view as shown in Figure \ref{fig:magneto}(b).

\begin{figure*}
\begin{interactive}{animation}{fig2_animation.mp4}
\centering
\includegraphics[width=16cm]{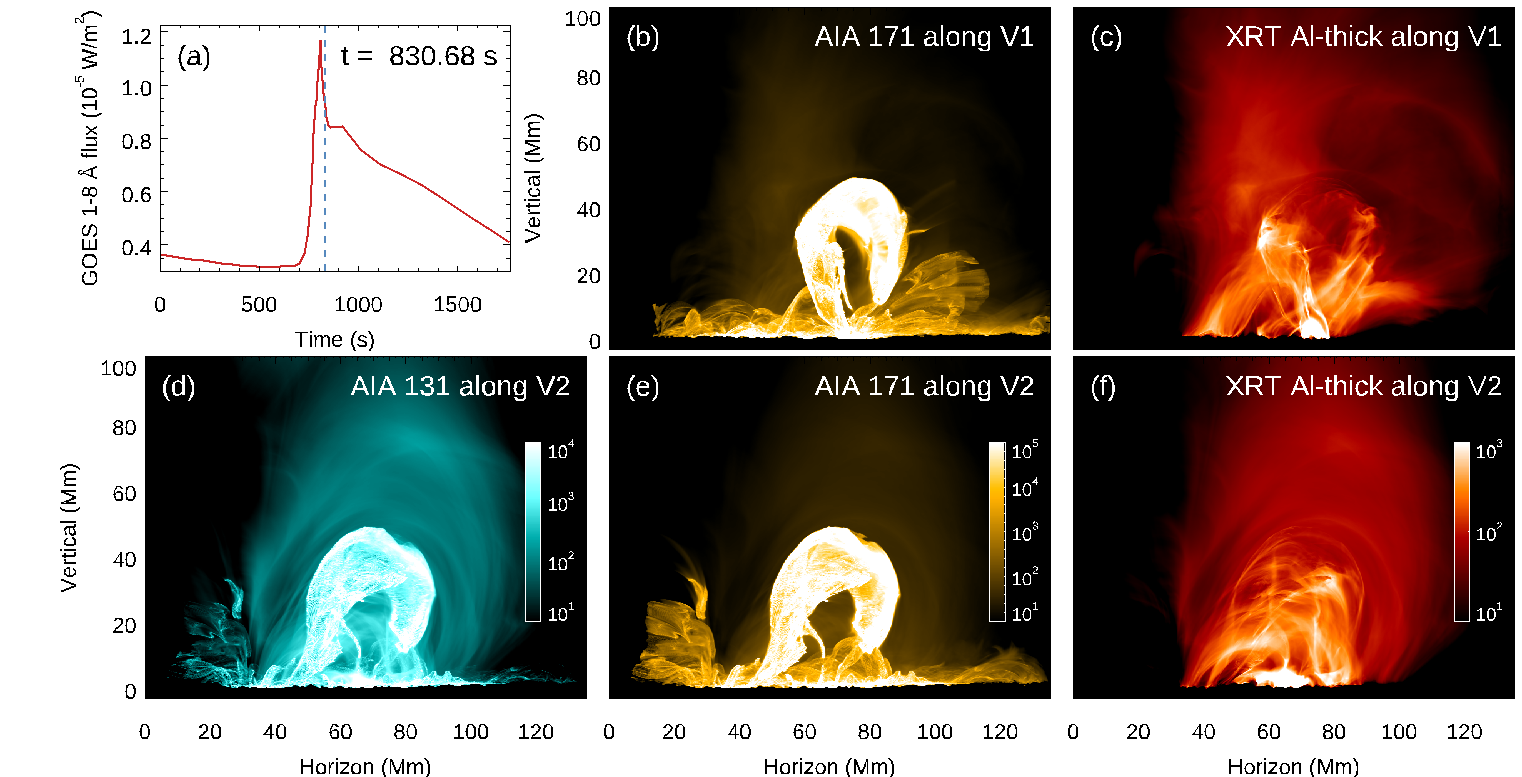} 
\end{interactive}
\caption{Synthesized observables of the event in the simulation. (a) Synthetic GOES 1–8\,\AA~flux in the selected region. The blue dotted line indicates the time instance ($t$ = 830.68 s) of the image shown in (b) -- (f). (b) Synthetic AIA 171\,\AA~image seen along V1 showing a rising bright arch. (c) Synthetic XRT Al-thick image seen along V1, showing a bright semicircular structure under the arch in (c). (d) Synthetic AIA 131\,\AA~image seen along V2. (e) Similar to (b) but along V2. (f) Similar to (c) but along V2. The animated version shows an evolution of 1764 s.
\newline(An animation of this figure is avaliable.)}\label{fig:obs}
\end{figure*}

To provide a global evolution of the flare, we display the synthetic {\it Geostationary Operational Environmental Satellites} (GOES) 1--8\,\AA~light curve in Figure \ref{fig:obs}(a). 
Since there is a relatively strong pre-eruption background, we use the increase of the synthetic GOES flux induced by the eruption to evaluate its flare class. In this event, the GOES flux increases by $8.5\times10^{-6}$\,W m$^{-2}$, and hence it corresponds to a C8.5 class flare. The synthetic GOES flux shows an impulsive rise phase followed by a slow decay phase, as was often found in observations. 

For a more intuitive comparison with multi-wavelength observations that are extensively used to diagnose plasma thermodynamics, we also synthesize the images of {\it Atmospheric Imaging Assembly} \citep[AIA,][]{AIA} 171\,\AA~and 131\,\AA~channels and {\it X-Ray Telescope} \citep[XRT,][]{XRT} Al-thick filter images that highlight plasma at different temperatures, respectively. To to better illustrate the key emission features during the eruption, we choose two side views as marked in Figure \ref{fig:magneto}(b): V1 is almost along the polarity inversion line (PIL) between P1 and N2, and V2 is perpendicular to V1. Figure \ref{fig:obs}(b) and (c) displays the synthetic images seen along V1 shortly after the peak time of the flare, and (d)--(f) seen along V2 at the same time. The associated animation shows the whole evolution of the eruption.

As shown in the animation of Figure \ref{fig:obs}, the first signature of the eruption appears at $t = 704.03$\,s as a bright blob at the bottom part of the synthetic EUV images below 10 Mm. Meanwhile it appears as an X-shaped compact structure in the XRT Al-thick images along V1, indicating that a small portion of plasma has been heated to relatively high temperatures in this early stage.

The bright blob seen in the EUV emission quickly rises and expands during the short impulsive phase. Shortly after the flare peak, the eruption appears as a bright arch in AIA 171 and 131\,\AA~images (Figure \ref{fig:obs}(b), (d) and (e)). We notice that the rising arch is impeded by the overlying loops and gradually slows down, as shown in Figure \ref{fig:obs}(d). At $t  =  858.89$\,s, only approximately 30 s after the time instant of Figure \ref{fig:obs}, the arch reaches its peak height, which is about 55\,Mm above the photosphere. Then, it breaks down and expands greatly, and eventually falls back to the solar surface. This eruption drives an EUV wave and subsequent quasi-periodic wave trains, which have been analysed in detail in \citet{Can2021}. 

The eruption appears significantly different in the synthetic soft X-ray images. The bright arch, which is the most prominent features seen in EUV emission, is absent in the XRT images. This suggests that the plasma confined in the arch may be heated to a few MK, but not significantly above 10\,MK. The XRT images show some thin threads that seem to outline the boundary of the EUV arch and a compact bright semicircular loop that lies below 10 Mm (Figure \ref{fig:obs}(c)). The image along the perspective of V2 (Figure \ref{fig:obs}(f)) reveals that the semicircular loop is the projection of a series of loops that lay along the perspective of V1. In the animation, we find a strong temporal correlation between the appearance of this semicircular loop and the impulsive phase of the GOES flux. With the flare going on, this loop becomes diffusive and evolves to a cusp-like shape that resembles post-flare loops in observed flares. Moreover, we notice a thin ray-like structure standing right above the semicircular loop. It outlines the current sheet in the eruption, as described later in this Letter.

\subsection{Morphology of the Magnetic Field\\ and the Current System}

\begin{figure*}
\centering
\includegraphics[width=\textwidth]{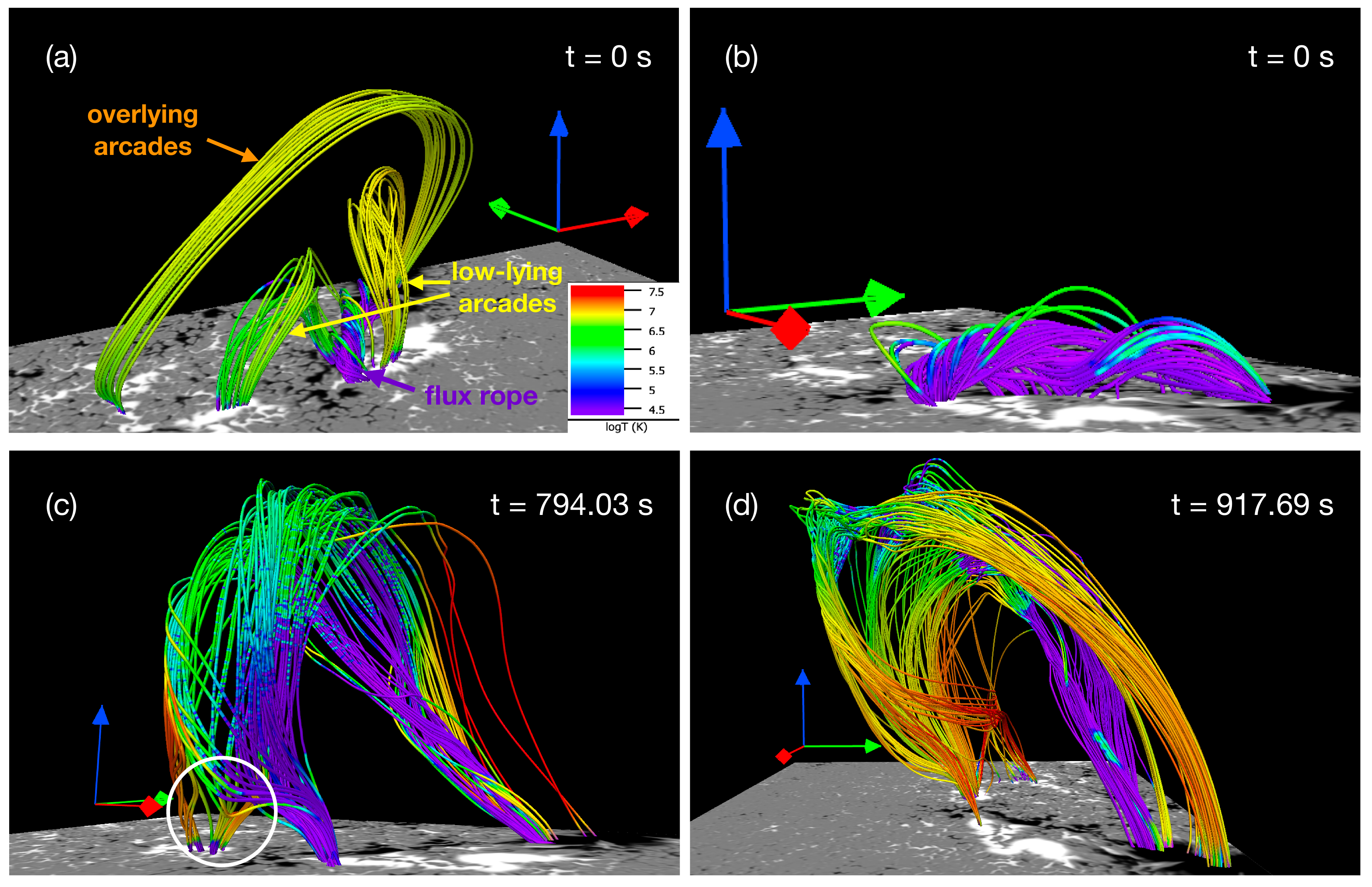}
\caption{Evolution of the magnetic flux rope before and during the eruption. (a) Coronal magnetic field configuration at $t_{0}$: a magnetic flux rope (purple arrow), two groups of low-lying arcades (yellow arrows) on both sides of the flux rope, and a group of large-scale overlying arcades (orange arrow) above. The colors of the field lines indicate the temperature of the local plasma, with the colorbar shown at the bottom right corner. The red, green, and blue arrows refer to the $x$-, $y$-, and $z$-axes in the simulation, respectively. (b)--(d) Magnetic field of the flux rope at different time instants. The white circle in (c) marks the position of flux transfer at the footpoint during the eruption. The 3D visualization is produced by VAPOR \citep{vapor2019}. }\label{fig:fr}
\end{figure*}

\begin{figure*}
\centering
\includegraphics[width=\textwidth]{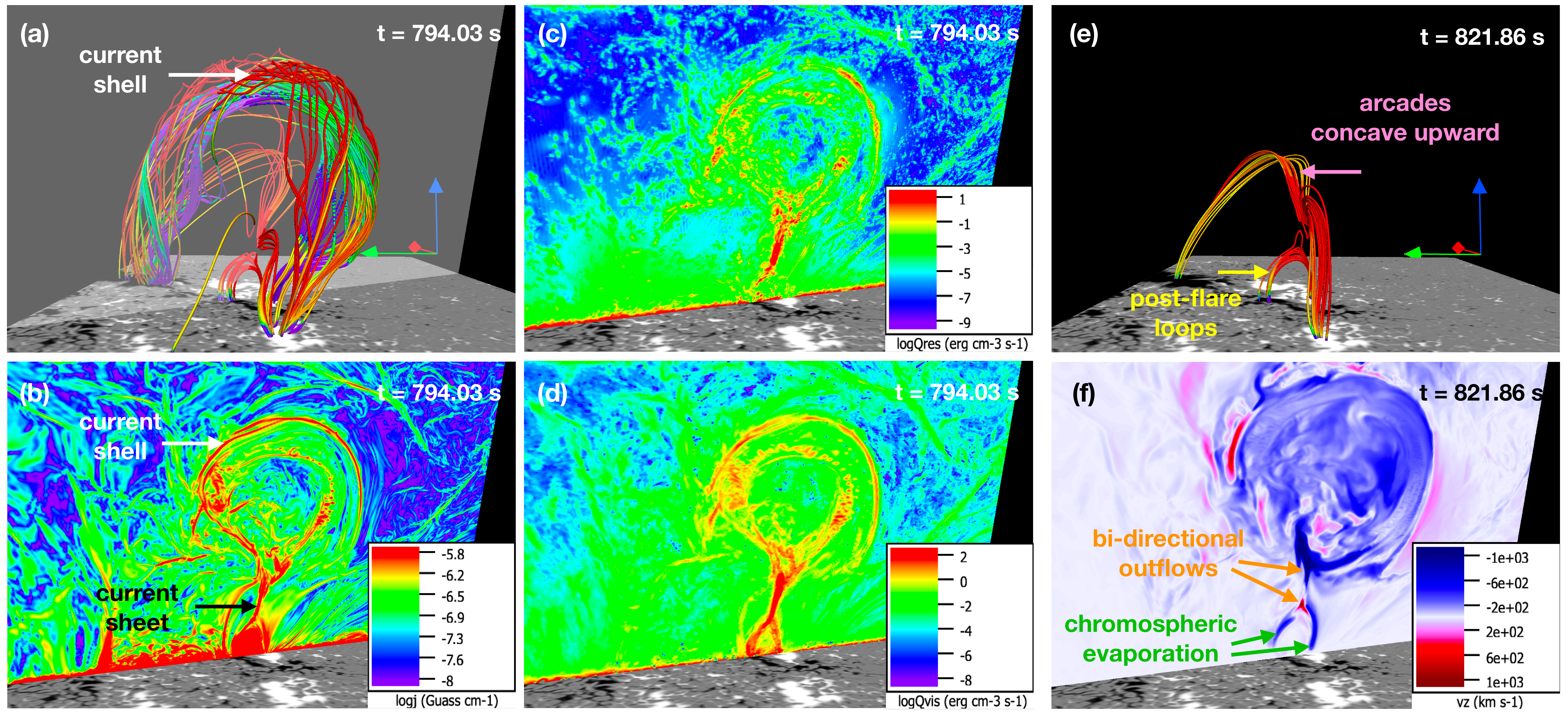}
\caption{Physical quantities related to the flux rope. (a) Overview of the magnetic field lines at $t$ = 794.03 s. The green and purple field lines show the same flux rope as that in Figure \ref{fig:fr}(d) but in another perspective, while the field lines within the current shell are plotted in red and indicated by white arrow. We select a slice (the transparent one in the image) to show the main physical parameters in the following panels. (b) Distribution of current density on the selected slice at the same time as in (a). The black arrow indicates the current sheet below the flux rope, and the white arrow marks the current shell. (c) Distribution of \rrr{\bf resistive} dissipation on the selected slice at 794.03 s. (d) Distribution of viscous dissipation on the selected slice at 794.03 s. (e) A highlight of field lines corresponding to the arcades concave upward (pink arrow) and post-flare loops (yellow arrow) formed after reconnection. (f) Distribution of plasma velocity on the selected slice at the same time as in (e), showing the high-speed bi-directional outflows (orange arrows) at both ends of the current sheet and the chromospheric evaporation (green arrows) in the post-flare loops. The 3D visualization is produced by VAPOR. }\label{fig:current}
\end{figure*}

The ejection of a coherent structure from the lower atmosphere, as described above, hints at the presence of a magnetic flux rope in this event. We calculate magnetic field lines based on seed points that are randomly distributed in selected regions. Figure \ref{fig:fr} presents a 3D view of the magnetic field lines. We note that, in order to highlight the evolution of the magnetic flux rope during different stages, the region of seed points varies for different panels in Figure \ref{fig:fr}.

As shown in Figure \ref{fig:fr}(a), the magnetic field at $t_{0}$ in the corona can be divided into three main parts: a pre-existing magnetic flux rope (purple arrow) \rrr{\bf of an average twist number of about $-$1.6} sitting above the PIL, a pair of low-lying arcades (yellow arrows) located on both sides of the flux rope, and a group of overlying arcades (orange arrows) that straddle over the flux rope. \rrr{\bf The flux rope was formed about 20 minute before $t_{0}$ through gradual reconnection between serpentine magnetic field lines along the PIL. Considering that the flux rope shows no obvious changes in morphology during this time period, we thus do not include the early evolution of the flux rope in this study.}

Three key snapshots during the evolution of the flux rope are shown in Figure \ref{fig:fr}(b)-(d). Figure \ref{fig:fr}(b) presents a zoomed view of the pre-existing flux rope, which consists of a branch of twisted field lines with their footpoints anchored in the major positive (negative) polarity P1 (N1). In the period we concern about, the flux rope evolves through three main stages: before the onset of the flare, the flux rope rises quasi-statically without obvious change in morphology; then, the flux rope erupts rapidly with expanding and rotating motions in the corona; finally, the flux rope slows down, stops at a height of approximately 55 Mm, and falls back to the solar surface, failing to develop into a CME. 

\rrr{\bf The dynamics the flux rope, in particular its relation to detailed force balance, is not the focus of the study presented in this Letter. Nevertheless, it is interesting to note here that torus instability \citep{Kliem2006} plays an important role in the initiation of the eruption, as the flux rope gradually rises to a height where the decay index of transverse component of the background magnetic field becomes sufficiently large. However, why the eruption is eventually confined can not be explained solely by the effect of torus instability, because both the flux rope and background field drastically change and interact during the eruption. The poloidal magnetic field of the flux rope is significantly changed in the latter evolution, such that the body of the flux rope presents a net downward Lorentz force. This is combined with a downward force exerted by the overlying arcades (as indicated in Figure \ref{fig:fr}(a)), which the flux rope has been pushing against during the eruption.

It is suggested that the topology of the flux rope may play an important role in determining its kinematic behaviours. For example, \citet{Zhong2021NC} demonstrated that the Lorentz force resulting from the non-axisymmetry of the flux rope can essentially constrain its eruption. The flux rope in our simulation exhibits a clearly more complex morphology than the simplified pre-eruption magnetic field structures in previous models. A thorough analysis on the kinematics of the eruption and driving forces will be presented in a forthcoming paper.}

Moreover, by comparing Figure \ref{fig:fr}(b) and (c), one can find a partial change of the footpoint of the flux rope: at the beginning, the positive footpoint of the flux rope is concentrated in P1; during the eruption, however, the flux rope bifurcates, such that part of the footpoint transfers to P2, as marked by the white circle in Figure \ref{fig:fr}(c). This phenomenon indicates a magnetic reconnection between the flux rope and the nearby arcades, which is similar to the process proposed by \citet{Aulanier2019} based on their simulations and is also found in observations \citep{Dudik2019,Xing2020}.

In Figure \ref{fig:current}, we plot several groups of magnetic field lines involved in the erupting flux rope and the distribution of some important physical quantities on a slice cutting through the cross-section of the flux rope. Figure \ref{fig:current}(b) displays the complex current system formed around the flux rope during the eruption. In particular, there is a circular region of strong current that wraps the flux rope, which looks like a `current shell' (white arrow in Figure \ref{fig:current}(b)). This current shell is co-spatial with the interface between the flux rope and the overlying arcades, namely a magnetic separatrix. As indicated by the white arrow in Panel (a), the magnetic field lines in the current shell are in corrugated shape, which implies the presence of strong magnetic disturbances in this region. 

Meanwhile, a thin current sheet is formed under the erupting flux rope, as indicated by the black arrow in Figure \ref{fig:current}(b). The low-lying arcades next to the flux rope (see Figure \ref{fig:fr}(a)) reconnect within the current sheet and transform to a branch of field lines concave upward (pink arrow in Panel (e)) and a group of cusp-shaped post-flare loops (yellow arrow in Panel (e)). \rrr{\bf The magnetic reconnection produces an upward Lorentz force that contributes to the rise of the flux rope, thus giving positive feedback to the eruption.}

\citet{Aulanier2019} described the reconnection between a pair of low-lying arcades, which results in a flux rope and a flare loop (termed as \emph{aa-rf}). The reconnection between the low-lying arcades in our simulation behaves slightly differently. We find that, at the initial time, the footpoints of the reconnecting arcades (P2 and N2) belong to the different magnetic domains with the footpoints of the flux rope (P1 and N1). Even though part of one footpoint of the flux rope is transferred from P1 to P2 during the eruption, the newly reconnected field lines neither show obvious twists as the flux rope, nor effectively supply magnetic flux to the flux rope.

We display the vertical velocity in the slice in Figure \ref{fig:current}(f). At the two ends of the current sheet, we find bi-directional outflows (orange arrows) with a high speed of more than 1000 km s$^{-1}$. There are also evident upflows along both legs of the post-flare loops (green arrows), which corresponds to chromospheric evaporation commonly seen in flare observations.

\subsection{Thermal Properties and Heating of the Plasma\\ Related to the Flux Rope}
\begin{figure*}
\centering
\includegraphics[width=\textwidth]{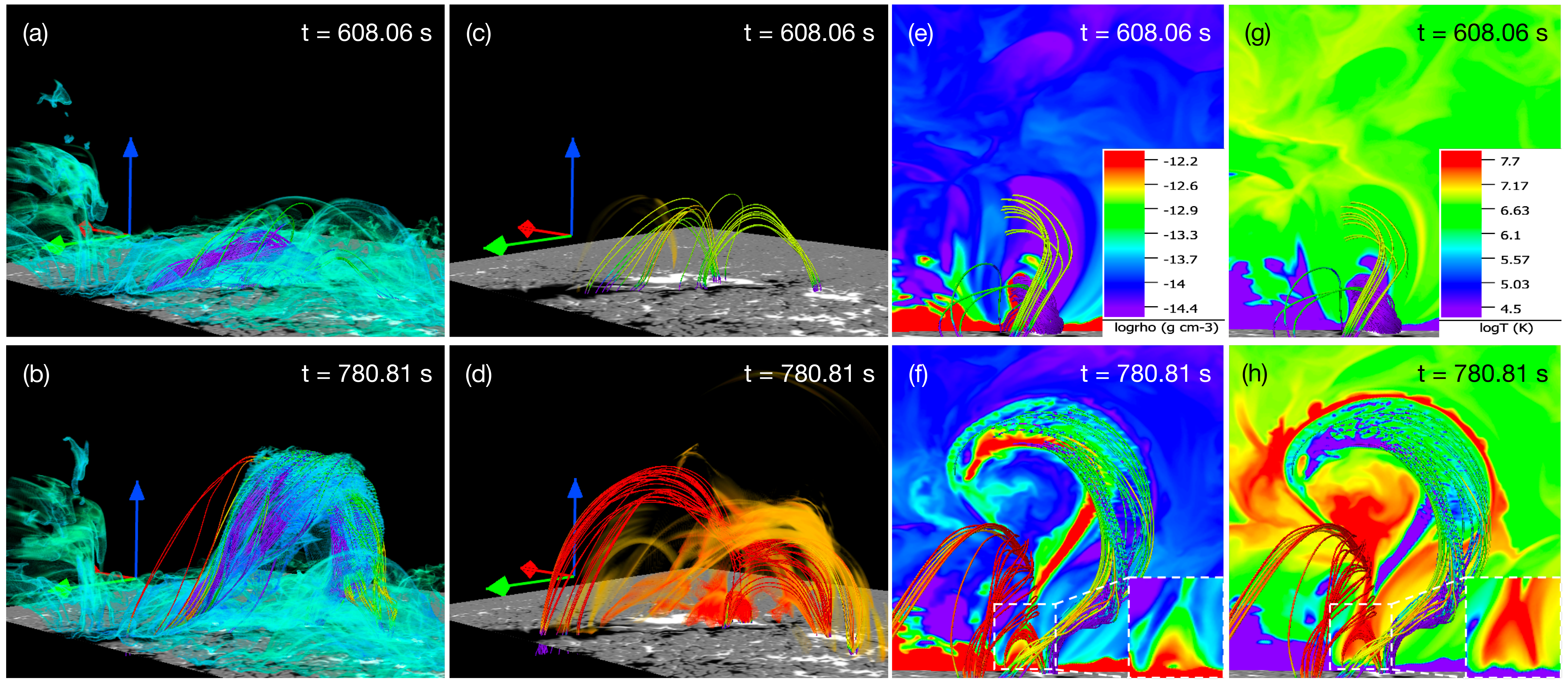}
\caption{Distribution of plasma parameters at two time instants of $t$ = 608.06 s (top row) and $t$ = 780.81 s (bottom row). Overplotted are selected magnetic field lines characterizing the event (flux rope in purple, low-lying arcades in yellow, and reconnected lines in red). (a) 3D distribution of the EM in the temperature range of 0.03--1 MK (green-colored structure) at $t$ = 608.06 s. (b)  Similar to (a) but at $t$ = 780.81 s. (c) 3D distribution of the EM in the temperature range of more than 10 MK (orange-colored structure) at $t$ = 608.06 s. (d)  Similar to (c) but at $t$ = 780.81 s. (e) Distribution of plasma density on the selected slice shown in Fig \ref{fig:current}(a) at $t$ = 608.06 s. (f) Similar to (e) but at $t$ = 780.81 s. The inset in (f) shows an enlarged view of the density within the post-flare loops. (g) Distribution of plasma temperature on the selected slice shown in Fig \ref{fig:current}(a) at $t$ = 608.06 s. (h) Similar to (g) but at $t$ = 780.81 s. The inset in (h) shows an enlarged view of the temperature within the post-flare loops. The 3D visualization is produced by VAPOR.}\label{fig:para}
\end{figure*}

To further relate the magnetic structures with the emission features of the event, we quantitatively analyse the main physical parameters of the plasma associated with the flux rope in two snapshots before ($t$ = 608.06 s) and during the impulsive phase of the eruption ($t$ = 780.81 s). 

Differential emission measures (DEM) deduced from multi-wavelength observations are extensively used to diagnose the temperature distribution of emitting plasma. However, the results suffer from the line-of-sight effect that may bring great uncertainties. With the simulation data, we evaluate the emission measure (EM) in the 3D space by
\begin{equation}
    {\rm EM}(x,y,z) = \sum \limits^{T}n_{e}^2(x,y,z,T)M(T),
\end{equation}
where $n_{e}$ is the electron number density and $M(T)$ is mask function of temperature $T$. For the analysis in this Letter, we use a low temperature mask: $M(T)=1$ if $4.5 \leq \log_{10}T \le 6.0$ and a high temperature mask: $M(T)=1$ if $\log_{10} T \geq 7$.  The 3D distributions of EMs in the low and high temperature ranges are shown in the first and second columns of Figure \ref{fig:para}, respectively. 

Before the eruption, the flux rope is embedded in cool plasma (Figure \ref{fig:para}(a)), and hot plasma is almost absent at this moment (Figure \ref{fig:para}(c)). At $t\approx781$ s, the flux rope is magnetic arch carrying plasma cooler than 1 MK (Figure \ref{fig:para}(b)), which is consistent with the results seen in synthetic observations shown in Figure \ref{fig:obs}. Meanwhile abundant plasma at temperatures higher than 10 MK can be found in the cusp-shaped post-flare loops and the magnetic arcades above, i.e., the magnetic field lines that involved in the reconnection triggered below the erupted flux rope.

The plasma density in the same slice as in Figure \ref{fig:current}(a) is shown in Figure \ref{fig:para}(e) and (f). It is seen that, before the eruption, the density of plasma hosted in the flux rope is more than two orders of magnitude higher than that of the background corona. During the eruption, the density decreases rapidly due to the expansion of the flux rope, but it remains significantly higher than the surrounding corona. As demonstrated by the insert in Figure \ref{fig:para}(f), the density in the post-flare loops is also higher than that outside the loops, as a result of the strong evaporation flows that feed plasma into the loops.

Figure \ref{fig:para}(g) and (h) shows the plasma temperature on the same slice. It clearly shows that before the eruption the temperature inside the flux rope is below 30000 K, two orders of magnitude lower than the temperature of the background corona. The high density and low temperature property of the flux rope indicates that the plasma hosted in it comes from the lower atmosphere.

As the flux rope erupts, a lot of hosted plasma is heated to temperatures of around 1 MK, giving rise to bright emission in the synthetic AIA images. Meanwhile, the plasma in the post-flare loops become hotter than 20 MK, which are the primary source of the soft X-ray emission seen in the synthetic XRT images and the GOES light curve. However, although being heated to high temperatures at the same time, the current shell wrapping around the flux rope is thin and of a low density, and hence can not yield a bright emission signature in the XRT images.

To understand the heating process during the eruption, we investigate the main heating sources in our simulation: resistive heating ($Q_{\rm res}$) and viscous heating ($Q_{\rm vis}$), which measure the dissipation of magnetic and kinetic energies to internal energy of plasma, respectively. The distribution of $Q_{\rm res}$ in the selected slice is shown in Figure \ref{fig:current}(c), and that for $Q_{\rm vis}$ is displayed in Figure \ref{fig:current}(d). Both $Q_{\rm res}$ and $Q_{\rm vis}$ increase greatly within the flux rope, providing a significant energy to heat the hosted plasma. The heating is also strong in the current sheet and current shell. The strong heating coincide with a lower plasma density in these regions. Therefore the temperatures efficiently rise to extremely high values.

We also notice that there are layered distributions of plasma inside the post-flare loops, which are shown in the insets of Figure \ref{fig:para}(f) and (h), implying sequential heating and cooling processes from inner to outer loops. The outer loops are newly formed and heated, with a higher temperature and a lower density, while the inner loops have already cooled down, with a lower temperature and a higher density. Similar results have been obtained in 2D and 2.5D simulations of solar flares \citep{Yokoyama2001,Takasao2015,Ye2020,Wang2021}.

\section{Summary and Conclusion} \label{sec:conclusion}
To explore the emission properties of flux ropes during their eruptions, we analyse a flux rope eruption associated with a C8.5 flare in a comprehensive 3D RMHD simulation. We focus on the topology of magnetic field and the thermal properties of the plasma involved with the flux rope and reach the main results as follows:
\begin{enumerate}
\item The initial coronal magnetic field consists of three main parts: a flux rope characterized by a branch of twisted field lines, a pair of low-lying arcades, and a group of overlying arcades. The flux rope has existed quasi-statically for a long time before its eruption. The eruption of the flux rope give rise to a C-class flare, however, itself fails to develop into a CME.

\item During the eruption of the flux rope, violent magnetic reconnection between the low-lying arcades occurs at the current sheet under it, generating a group of post-flare loops. There appears a current shell wrapping the flux rope, which may result from the squeezing between the flux rope and the overlying field. 

\item The plasma within the flux rope is initially much denser and cooler than that in the surrounding corona, and it can be heated to around 1 MK during the eruption. The plasmas with the highest temperature of more than 10 MK mainly appear at the current sheet below the flux loop, the current shell around it, and the post-flare loops.

\item  The pre-existing flux rope is hardly visible in the AIA 171 \AA~images before its eruption but manifested as a bright arch during its eruption. The current sheet and post-flare loops can be clearly identified in the synthetic XRT images.

\end{enumerate}

When the flux rope is formed is a key question in the study on solar eruptions. Previous theoretical models have suggested two plausible approaches: one indicates that the pre-existing flux ropes is an important prerequisite for solar eruptions \citep{Fan2017,Aulanier2010}, while the other argues that the pre-eruption configuration can be sheared magnetic arcades and the flux ropes are products of the eruptions \citep{Jiang2021}. Observations also lead to diverse results on the formation of flux ropes. The hot channel structure, which usually shows up in AIA 94\,\AA~and 131\,\AA~images that are sensitive to hot plasma, is regarded as a credible candidate of the flux ropes. Most of the hot channels show their earliest signatures before the onset of eruptions \citep{Zhang2012,Cheng2013}. However, there are also events with hot channels forming during the impulsive acceleration phase of the CME \citep{cheng2011}. 

Interestingly, we notice in our simulation that there are no significant emission features of the pre-existing flux rope in the synthetic AIA images before the flare onset. Whether it is visible in other cooler wavelengths like H$\alpha$ requires further investigations. Our results also indicates that during the eruption, the emission features in low or high temperatures can only exhibit limited information of the magnetic field. The arch-like structure seen in cool EUV channels (e.g, AIA 171) outlines the main body of the flux rope, and the hot channels may correspond to only the significantly heated part of the flux ropes, such as the concave arcades formed by the reconnection below the erupted flux rope. To conclude, the flux rope and involved plasma in our simulation is a multi-thermal structure and their observational properties change drastically. Therefore, even in events without hot channels before the eruption, we cannot preclude the possibility of a pre-existing flux rope. More comprehensive criteria need to be considered when judging the formation of the flux ropes.

\begin{acknowledgments}
\rrr{\bf The authors thanks the anonymous referee for helpful suggestions that improve the clarity of this Letter.} We are grateful to Xin Cheng and Chen Xing for inspiring discussions. This work was supported by National Key R\&D Program of China under grant 2021YFA1600504 and by NSFC under grants 11733003 and 12127901. F.C. acknowledges the support from the Program for Innovative Talents and Entrepreneurs in Jiangsu. This material is based upon work supported by the National Center for Atmospheric Research, which is a major facility sponsored by the National Science Foundation under Cooperative Agreement No. 1852977. The high-performance computing support is provided by Cheyenne (doi:10.5065/D6RX99HX). F.C. was supported by the Advanced Study Program postdoctoral fellowship at NCAR and by the George Ellery Hale postdoctoral fellowship at the University of Colorado Boulder.
\end{acknowledgments}

\bibliography{manuscript}{}

\begin{thebibliography}{}
\expandafter\ifx\csname natexlab\endcsname\relax\def\natexlab#1{#1}\fi
\providecommand{\url}[1]{\href{#1}{#1}}
\providecommand{\dodoi}[1]{doi:~\href{http://doi.org/#1}{\nolinkurl{#1}}}
\providecommand{\doeprint}[1]{\href{http://ascl.net/#1}{\nolinkurl{http://ascl.net/#1}}}
\providecommand{\doarXiv}[1]{\href{https://arxiv.org/abs/#1}{\nolinkurl{https://arxiv.org/abs/#1}}}

\bibitem[{{Amari} {et~al.}(2003){Amari}, {Luciani}, {Aly}, {Mikic}, \&
  {Linker}}]{Amari2003}
{Amari}, T., {Luciani}, J.~F., {Aly}, J.~J., {Mikic}, Z., \& {Linker}, J. 2003,
  \apj, 585, 1073, \dodoi{10.1086/345501}

\bibitem[{{Amari} {et~al.}(2000){Amari}, {Luciani}, {Mikic}, \&
  {Linker}}]{Amari2000}
{Amari}, T., {Luciani}, J.~F., {Mikic}, Z., \& {Linker}, J. 2000, \apjl, 529,
  L49, \dodoi{10.1086/312444}

\bibitem[{{Aulanier} \& {Dud{\'\i}k}(2019)}]{Aulanier2019}
{Aulanier}, G., \& {Dud{\'\i}k}, J. 2019, \aap, 621, A72,
  \dodoi{10.1051/0004-6361/201834221}

\bibitem[{{Aulanier} {et~al.}(2010){Aulanier}, {T{\"o}r{\"o}k}, {D{\'e}moulin},
  \& {DeLuca}}]{Aulanier2010}
{Aulanier}, G., {T{\"o}r{\"o}k}, T., {D{\'e}moulin}, P., \& {DeLuca}, E.~E.
  2010, \apj, 708, 314, \dodoi{10.1088/0004-637X/708/1/314}

\bibitem[{{Chen} {et~al.}(2021){Chen}, {Rempel}, \& {Fan}}]{Chen2021}
{Chen}, F., {Rempel}, M., \& {Fan}, Y. 2021, arXiv e-prints, arXiv:2106.14055.
\newblock \doarXiv{2106.14055}

\bibitem[{{Cheng} {et~al.}(2017){Cheng}, {Guo}, \& {Ding}}]{Cheng2017}
{Cheng}, X., {Guo}, Y., \& {Ding}, M. 2017, Science China Earth Sciences, 60,
  1383, \dodoi{10.1007/s11430-017-9074-6}

\bibitem[{{Cheng} {et~al.}(2013){Cheng}, {Zhang}, {Ding}, {Liu}, \&
  {Poomvises}}]{Cheng2013}
{Cheng}, X., {Zhang}, J., {Ding}, M.~D., {Liu}, Y., \& {Poomvises}, W. 2013,
  \apj, 763, 43, \dodoi{10.1088/0004-637X/763/1/43}

\bibitem[{{Cheng} {et~al.}(2020){Cheng}, {Zhang}, {Kliem}, {T{\"o}r{\"o}k},
  {Xing}, {Zhou}, {Inhester}, \& {Ding}}]{Cheng2020}
{Cheng}, X., {Zhang}, J., {Kliem}, B., {et~al.} 2020, \apj, 894, 85,
  \dodoi{10.3847/1538-4357/ab886a}

\bibitem[{{Cheng} {et~al.}(2011){Cheng}, {Zhang}, {Liu}, \& {Ding}}]{cheng2011}
{Cheng}, X., {Zhang}, J., {Liu}, Y., \& {Ding}, M.~D. 2011, \apjl, 732, L25,
  \dodoi{10.1088/2041-8205/732/2/L25}

\bibitem[{{Cheung} {et~al.}(2019){Cheung}, {Rempel}, {Chintzoglou}, {Chen},
  {Testa}, {Mart{\'\i}nez-Sykora}, {Sainz Dalda}, {DeRosa}, {Malanushenko},
  {Hansteen}, {De Pontieu}, {Carlsson}, {Gudiksen}, \& {McIntosh}}]{Cheung2019}
{Cheung}, M.~C.~M., {Rempel}, M., {Chintzoglou}, G., {et~al.} 2019, Nature
  Astronomy, 3, 160, \dodoi{10.1038/s41550-018-0629-3}

\bibitem[{{Dud{\'\i}k} {et~al.}(2019){Dud{\'\i}k}, {L{\"o}rin{\v{c}}{\'\i}k},
  {Aulanier}, {Zemanov{\'a}}, \& {Schmieder}}]{Dudik2019}
{Dud{\'\i}k}, J., {L{\"o}rin{\v{c}}{\'\i}k}, J., {Aulanier}, G.,
  {Zemanov{\'a}}, A., \& {Schmieder}, B. 2019, \apj, 887, 71,
  \dodoi{10.3847/1538-4357/ab4f86}

\bibitem[{{Fan}(2017)}]{Fan2017}
{Fan}, Y. 2017, \apj, 844, 26, \dodoi{10.3847/1538-4357/aa7a56}

\bibitem[{{Fan} \& {Liu}(2019)}]{Fan2019}
{Fan}, Y., \& {Liu}, T. 2019, Frontiers in Astronomy and Space Sciences, 6, 27,
  \dodoi{10.3389/fspas.2019.00027}

\bibitem[{{Filippov} {et~al.}(2015){Filippov}, {Martsenyuk}, {Srivastava}, \&
  {Uddin}}]{Filippov2015}
{Filippov}, B., {Martsenyuk}, O., {Srivastava}, A.~K., \& {Uddin}, W. 2015,
  Journal of Astrophysics and Astronomy, 36, 157,
  \dodoi{10.1007/s12036-015-9321-5}

\bibitem[{{Gibson}(2015)}]{Sarah2015}
{Gibson}, S. 2015, in Astrophysics and Space Science Library, Vol. 415, Solar
  Prominences, ed. J.-C. {Vial} \& O.~{Engvold}, 323,
  \dodoi{10.1007/978-3-319-10416-4\_13}

\bibitem[{{Golub} {et~al.}(2007){Golub}, {Deluca}, {Austin}, {Bookbinder},
  {Caldwell}, {Cheimets}, {Cirtain}, {Cosmo}, {Reid}, {Sette}, {Weber},
  {Sakao}, {Kano}, {Shibasaki}, {Hara}, {Tsuneta}, {Kumagai}, {Tamura},
  {Shimojo}, {McCracken}, {Carpenter}, {Haight}, {Siler}, {Wright}, {Tucker},
  {Rutledge}, {Barbera}, {Peres}, \& {Varisco}}]{XRT}
{Golub}, L., {Deluca}, E., {Austin}, G., {et~al.} 2007, \solphys, 243, 63,
  \dodoi{10.1007/s11207-007-0182-1}

\bibitem[{{Green} {et~al.}(2007){Green}, {Kliem}, {T{\"o}r{\"o}k}, {van
  Driel-Gesztelyi}, \& {Attrill}}]{Green2007}
{Green}, L.~M., {Kliem}, B., {T{\"o}r{\"o}k}, T., {van Driel-Gesztelyi}, L., \&
  {Attrill}, G.~D.~R. 2007, \solphys, 246, 365,
  \dodoi{10.1007/s11207-007-9061-z}

\bibitem[{{Guo} {et~al.}(2019){Guo}, {Xia}, {Keppens}, {Ding}, \&
  {Chen}}]{Guo2019}
{Guo}, Y., {Xia}, C., {Keppens}, R., {Ding}, M.~D., \& {Chen}, P.~F. 2019,
  \apjl, 870, L21, \dodoi{10.3847/2041-8213/aafabf}

\bibitem[{{He} {et~al.}(2020){He}, {Jiang}, {Zou}, {Duan}, {Feng}, {Zuo}, \&
  {Wang}}]{He2020}
{He}, W., {Jiang}, C., {Zou}, P., {et~al.} 2020, \apj, 892, 9,
  \dodoi{10.3847/1538-4357/ab75ab}

\bibitem[{{Inoue} {et~al.}(2014){Inoue}, {Hayashi}, {Magara}, {Choe}, \&
  {Park}}]{Inoue2014}
{Inoue}, S., {Hayashi}, K., {Magara}, T., {Choe}, G.~S., \& {Park}, Y.~D. 2014,
  \apj, 788, 182, \dodoi{10.1088/0004-637X/788/2/182}

\bibitem[{{Jiang} {et~al.}(2021){Jiang}, {Feng}, {Liu}, {Yan}, {Hu}, {Moore},
  {Duan}, {Cui}, {Zuo}, {Wang}, \& {Wei}}]{Jiang2021}
{Jiang}, C., {Feng}, X., {Liu}, R., {et~al.} 2021, Nature Astronomy, 5, 1126,
  \dodoi{10.1038/s41550-021-01414-z}

\bibitem[{{Kliem} {et~al.}(2013){Kliem}, {Su}, {van Ballegooijen}, \&
  {DeLuca}}]{Kliem2013}
{Kliem}, B., {Su}, Y.~N., {van Ballegooijen}, A.~A., \& {DeLuca}, E.~E. 2013,
  \apj, 779, 129, \dodoi{10.1088/0004-637X/779/2/129}

\bibitem[{{Kliem} \& {T{\"o}r{\"o}k}(2006)}]{Kliem2006}
{Kliem}, B., \& {T{\"o}r{\"o}k}, T. 2006, \prl, 96, 255002,
  \dodoi{10.1103/PhysRevLett.96.255002}

\bibitem[{{Lemen} {et~al.}(2012){Lemen}, {Title}, {Akin}, {Boerner}, {Chou},
  {Drake}, {Duncan}, {Edwards}, {Friedlaender}, {Heyman}, {Hurlburt}, {Katz},
  {Kushner}, {Levay}, {Lindgren}, {Mathur}, {McFeaters}, {Mitchell}, {Rehse},
  {Schrijver}, {Springer}, {Stern}, {Tarbell}, {Wuelser}, {Wolfson}, {Yanari},
  {Bookbinder}, {Cheimets}, {Caldwell}, {Deluca}, {Gates}, {Golub}, {Park},
  {Podgorski}, {Bush}, {Scherrer}, {Gummin}, {Smith}, {Auker}, {Jerram},
  {Pool}, {Soufli}, {Windt}, {Beardsley}, {Clapp}, {Lang}, \& {Waltham}}]{AIA}
{Lemen}, J.~R., {Title}, A.~M., {Akin}, D.~J., {et~al.} 2012, \solphys, 275,
  17, \dodoi{10.1007/s11207-011-9776-8}

\bibitem[{Li {et~al.}(2019)Li, Jaroszynski, Pearse, Orf, \& Clyne}]{vapor2019}
Li, S., Jaroszynski, S., Pearse, S., Orf, L., \& Clyne, J. 2019, Atmosphere,
  10, \dodoi{10.3390/atmos10090488}

\bibitem[{{Lin}(2004)}]{Lin2004}
{Lin}, J. 2004, \solphys, 219, 169, \dodoi{10.1023/B:SOLA.0000021798.46677.16}

\bibitem[{{Liu}(2020)}]{Liu2020}
{Liu}, R. 2020, Research in Astronomy and Astrophysics, 20, 165,
  \dodoi{10.1088/1674-4527/20/10/165}

\bibitem[{{Mackay} {et~al.}(2010){Mackay}, {Karpen}, {Ballester}, {Schmieder},
  \& {Aulanier}}]{Mackay2010}
{Mackay}, D.~H., {Karpen}, J.~T., {Ballester}, J.~L., {Schmieder}, B., \&
  {Aulanier}, G. 2010, \ssr, 151, 333, \dodoi{10.1007/s11214-010-9628-0}

\bibitem[{{McKenzie} \& {Canfield}(2008)}]{McKenzie2008}
{McKenzie}, D.~E., \& {Canfield}, R.~C. 2008, \aap, 481, L65,
  \dodoi{10.1051/0004-6361:20079035}

\bibitem[{{Nindos} {et~al.}(2015){Nindos}, {Patsourakos}, {Vourlidas}, \&
  {Tagikas}}]{Nindos2015}
{Nindos}, A., {Patsourakos}, S., {Vourlidas}, A., \& {Tagikas}, C. 2015, \apj,
  808, 117, \dodoi{10.1088/0004-637X/808/2/117}

\bibitem[{{Ouyang} {et~al.}(2017){Ouyang}, {Zhou}, {Chen}, \&
  {Fang}}]{Ouyang2017}
{Ouyang}, Y., {Zhou}, Y.~H., {Chen}, P.~F., \& {Fang}, C. 2017, \apj, 835, 94,
  \dodoi{10.3847/1538-4357/835/1/94}

\bibitem[{{Patsourakos} {et~al.}(2020){Patsourakos}, {Vourlidas},
  {T{\"o}r{\"o}k}, {Kliem}, {Antiochos}, {Archontis}, {Aulanier}, {Cheng},
  {Chintzoglou}, {Georgoulis}, {Green}, {Leake}, {Moore}, {Nindos}, {Syntelis},
  {Yardley}, {Yurchyshyn}, \& {Zhang}}]{Patsourakos2020}
{Patsourakos}, S., {Vourlidas}, A., {T{\"o}r{\"o}k}, T., {et~al.} 2020, \ssr,
  216, 131, \dodoi{10.1007/s11214-020-00757-9}

\bibitem[{{R{\'e}gnier} {et~al.}(2011){R{\'e}gnier}, {Walsh}, \&
  {Alexander}}]{Regnier2011}
{R{\'e}gnier}, S., {Walsh}, R.~W., \& {Alexander}, C.~E. 2011, \aap, 533, L1,
  \dodoi{10.1051/0004-6361/201117381}

\bibitem[{{Rempel}(2017)}]{Rempel:2017}
{Rempel}, M. 2017, \apj, 834, 10, \dodoi{10.3847/1538-4357/834/1/10}

\bibitem[{{Schmieder} {et~al.}(2015){Schmieder}, {Aulanier}, \&
  {Vr{\v{s}}nak}}]{Schmieder2015}
{Schmieder}, B., {Aulanier}, G., \& {Vr{\v{s}}nak}, B. 2015, \solphys, 290,
  3457, \dodoi{10.1007/s11207-015-0712-1}

\bibitem[{{Takasao} {et~al.}(2015){Takasao}, {Matsumoto}, {Nakamura}, \&
  {Shibata}}]{Takasao2015}
{Takasao}, S., {Matsumoto}, T., {Nakamura}, N., \& {Shibata}, K. 2015, \apj,
  805, 135, \dodoi{10.1088/0004-637X/805/2/135}

\bibitem[{{V{\"o}gler} {et~al.}(2005){V{\"o}gler}, {Shelyag}, {Sch{\"u}ssler},
  {Cattaneo}, {Emonet}, \& {Linde}}]{Voegler+al:2005}
{V{\"o}gler}, A., {Shelyag}, S., {Sch{\"u}ssler}, M., {et~al.} 2005, \aap, 429,
  335, \dodoi{10.1051/0004-6361:20041507}

\bibitem[{{Wang} {et~al.}(2021{\natexlab{a}}){Wang}, {Chen}, \&
  {Ding}}]{Can2021}
{Wang}, C., {Chen}, F., \& {Ding}, M. 2021{\natexlab{a}}, \apjl, 911, L8,
  \dodoi{10.3847/2041-8213/abefe6}

\bibitem[{{Wang} {et~al.}(2021{\natexlab{b}}){Wang}, {Cheng}, {Ding}, \&
  {Lu}}]{Wang2021}
{Wang}, Y., {Cheng}, X., {Ding}, M., \& {Lu}, Q. 2021{\natexlab{b}}, \apj, 923,
  227, \dodoi{10.3847/1538-4357/ac3142}

\bibitem[{{Webb} \& {Howard}(2012)}]{Webb2012}
{Webb}, D.~F., \& {Howard}, T.~A. 2012, Living Reviews in Solar Physics, 9, 3,
  \dodoi{10.12942/lrsp-2012-3}

\bibitem[{{Xia} {et~al.}(2014){Xia}, {Keppens}, {Antolin}, \&
  {Porth}}]{Xia2014}
{Xia}, C., {Keppens}, R., {Antolin}, P., \& {Porth}, O. 2014, \apjl, 792, L38,
  \dodoi{10.1088/2041-8205/792/2/L38}

\bibitem[{{Xing} {et~al.}(2020){Xing}, {Cheng}, \& {Ding}}]{Xing2020}
{Xing}, C., {Cheng}, X., \& {Ding}, M.~D. 2020, The Innovation, 1, 100059,
  \dodoi{10.1016/j.xinn.2020.100059}

\bibitem[{{Ye} {et~al.}(2020){Ye}, {Cai}, {Shen}, {Raymond}, {Lin}, {Roussev},
  \& {Mei}}]{Ye2020}
{Ye}, J., {Cai}, Q., {Shen}, C., {et~al.} 2020, \apj, 897, 64,
  \dodoi{10.3847/1538-4357/ab93b5}

\bibitem[{{Yokoyama} \& {Shibata}(2001)}]{Yokoyama2001}
{Yokoyama}, T., \& {Shibata}, K. 2001, \apj, 549, 1160, \dodoi{10.1086/319440}

\bibitem[{{Zhang} {et~al.}(2012){Zhang}, {Cheng}, \& {Ding}}]{Zhang2012}
{Zhang}, J., {Cheng}, X., \& {Ding}, M.-D. 2012, Nature Communications, 3, 747,
  \dodoi{10.1038/ncomms1753}

\bibitem[{{Zhang} {et~al.}(2001){Zhang}, {Dere}, {Howard}, {Kundu}, \&
  {White}}]{Zhang2001}
{Zhang}, J., {Dere}, K.~P., {Howard}, R.~A., {Kundu}, M.~R., \& {White}, S.~M.
  2001, \apj, 559, 452, \dodoi{10.1086/322405}

\bibitem[{{Zhong} {et~al.}(2021){Zhong}, {Guo}, \& {Ding}}]{Zhong2021NC}
{Zhong}, Z., {Guo}, Y., \& {Ding}, M.~D. 2021, Nature Communications, 12, 2734,
  \dodoi{10.1038/s41467-021-23037-8}

\end{thebibliography}
\bibliographystyle{aasjournal}

\listofchanges
\end{document}